\documentclass[twocolumn,prd,nofootinbib,aps,prl,floats,floatfix,amsmath,amssymb,longbibliography,secnumarabic]{revtex4-1} %
\usepackage[english]{babel}

\usepackage[letterpaper,top=2cm,bottom=2cm,left=3cm,right=3cm,marginparwidth=1.75cm]{geometry}

\usepackage{amsmath,cancel}
\usepackage{graphicx,verbatim}
\usepackage[colorlinks=true,citecolor=red]{hyperref}
\usepackage{xcolor}

\newcommand{\be}{\begin{equation}}
\newcommand{\ee}{\end{equation}}
\def\sfrac#1#2{{\textstyle{#1\over #2}}}
\newcommand{\bea}{\begin{eqnarray}}
\newcommand{\eea}{\end{eqnarray}}
\newcommand{\nn}{\nonumber}

\newcommand{\sss}{\scriptscriptstyle}

\begin{document}

\title{Quintessential dark energy crossing the phantom divide}
\author{Ruiqi Chen}
\author{James M.\ Cline}
\email{jcline@physics.mcgill.ca}
\author{Varun Muralidharan}
\email{varun.muralidharan@mail.mcgill.ca}
\author{Benjamin Salewicz}
\affiliation{McGill University Department of Physics \& Trottier Space Institute, 3600 Rue University, Montr\'eal, QC, H3A 2T8, Canada}

\begin{abstract}
Motivated by recent results from the DESI collaboration, 
we explore two classes of quintessence models that can give rise to crossing of the dark energy equation of state through the ``phantom divide'' $w=-1$.
These are models with Lagrangians that involve higher powers of the kinetic energy $\dot\phi^2$, or where the dark matter (DM) mass is a function of $\phi$.  Both have similar features with respect to the reconstructed 
redshift-dependent $w(z)$: moderate tuning of parameters is required to achieve the desired shape, and it is difficult or impossible for $w(z)$ to continue evolving smoothly as $z$ becomes large.  Nevertheless, they
give a strong improvement over $\Lambda$CDM in fitting the data.
We point out that models of coupled dark matter and dark energy that cross the phantom divide are under pressure from constraints on long-range DM forces.  They rule out the simplest renormalizable coupling of scalar DM to quintessence, but leave the fermionic case allowed, and exponentially coupled models of either kind of DM are safe from current constraints.
\end{abstract}

\maketitle

\section{Introduction}
The null energy condition (NEC) states that the stress-energy tensor should be nonnegative along any lightlike geodesic, $T_{\mu\nu}u^\mu u^\nu \ge 0$.  For a perfect fluid, this constrains the pressure and energy density to obey $w = p/\rho \ge -1$.  The relation is satisfied for known sources of energy density, and saturated by cosmological constant.  It is violated by exotic sources such as the Casimir effect \cite{Alexandre:2021imu}, or phantom
scalar fields that have a wrong-sign kinetic term
\cite{Caldwell:1999ew}.

Experimental motivation for violating the NEC arose in 2003 when WMAP's first data release suggested  $w < -1$ for the dark energy of the Universe \cite{WMAP:2003elm}, although $w=-1$ was also consistent with the data.  Some twenty years later,
the DESI collaboration published evidence for 
evolving dark energy \cite{DESI:2024aqx} which showed a preference for $w<-1$ at redshifts $z\gtrsim 0.5$.
More recently DESI has analyzed baryon acoustic oscillations from their second data release and given stronger evidence for evolving dark energy
\cite{DESI:2025fii,DESI:2025wyn,DESI:2025zgx}.
Unlike the WMAP data, which only tried to constrain the present value of $w$, these results fit the evolution of $w(z)$ back to $z\sim 3$, and continue to favor a crossing from $w<-1$ to $w>-1$ at $z\sim 0.5$.
A variety of complementary analyses \cite{Cheng:2025lod, Gonzalez-Fuentes:2025lei,Ye:2025ark,Keeley:2025rlg,Silva:2025twg,Ishak:2025cay,Ozulker:2025ehg,Chaudhary:2025vzy,Wang:2025vtw,RoyChoudhury:2025dhe} reinforce this conclusion; see however Refs.\ \cite{Verma:2025ujt,Roy:2025cxk,Gialamas:2025pwv}.

It is possible to have an apparent phantom crossing in a relatively mundane way, if quintessence couples to dark matter (DM) and causes the DM mass to evolve with time \cite{Das:2005yj}.
This has been shown to be sufficient to explain the observations \cite{Chakraborty:2025syu, Khoury:2025txd, Guedezounme:2025wav},
and is perhaps the most plausible mechanism if they continue to be upheld.    A related idea has been explored in Ref.\ \cite{Braglia:2025gdo}.

On the other hand, it is not easy to get a true phantom crossing, coming from an isolated dark energy sector.  Ordinary quintessence models have $w>-1$, so they do not provide an optimal fit to the DESI data \cite{Cline:2025sbt}. Even standard phantom models are not capable of crossing the phantom divide, although they can fall on the wrong side of it, as we will discuss below Eq.\ (\ref{wgenrel}).  Instead, we consider
generalizations in which the Lagrangian is a nonlinear function $f(X)$ of the quintessence kinetic term
$X = \sfrac12(\partial\phi)^2$, sometimes known as 
ghost condensate models \cite{Arkani-Hamed:2003pdi}.

Quintom models are another possible framework for crossing the phantom divide, where ordinary quintessence couples to a phantom field \cite{Guo:2004fq}.
Such models have been shown to give a good fit to the present data \cite{Gomez-Valent:2025mfl}.  
Alternatively, modified gravity theories can give rise to phantom crossing \cite{Hogas:2021fmr,Nojiri:2025low,Yao:2025wlx,Tsujikawa:2025xxx,Hogas:2025ahb},
as well as theories of quintessence nonminimally coupled to gravity \cite{jysf-k72m,PhysRevD.111.L041303,Ye:2024ywg,Wang:2025znm}.
In this paper we restrict our study to single-field,
minimally coupled quintessence models, excluding $k$-essence
(Lagrangians of the form $f(X)V(\phi)$), which was
recently considered in Ref.\ \cite{Goldstein:2025epp}.

In the following, we investigate two classes of models---ghost condensation and quintessence-DM couplings---with respect to their ability to reproduce a desired $w(z)$ history.  A striking similarity between the two models emerges: they both predict undulatory $w(z)$ curves that tend to become unstable at large $z$ when the equations of motion are integrated backwards into the past.  To make this instability appear only at large $z$, beyond which the current observations are constraining, tuning of the model parameters is required.  The qualitative shapes of the predicted $w(z)$ curves are similar, such that it would be difficult to distinguish the two kinds of models on this basis alone.

\section{Ghost condensate models}
We first consider quintessence with a nonlinear 
kinetic function in the Lagrangian,
\be
    {\cal L} = f(X) - V(\phi)\,,
\ee
where $X = \sfrac12(\partial\phi)^2$.  In this work we will neglect inhomogeneities (except for perturbations included in \texttt{Cobaya} \cite{Torrado:2020dgo}) and assume that
$\phi = \phi(t)$ only, so that $X = \sfrac12\dot\phi^2$.

In general, such theories are prone to instabilities at both the classical and quantum levels.  The quantum instability is due to the $\phi$ particles carrying negative energy, hence leading to spontaneous decay of the vacuum into 
$\phi\phi\gamma\gamma$ \cite{Carroll:2003st,Cline:2003gs}, whenever $f'(X)<0$.
One can make sense of such models by assuming they are low-energy effective theories, valid below a momentum cutoff $\Lambda_{\sss \rm UV} \sim 18\,$MeV
\cite{Cline:2023cwm}.   The classical instability occurs when $f' + 2Xf''\le 0$ \cite{Copeland:2006wr}, corresponding to negative squared sound speed.  Since the phantom crossing lasts only for a small interval of redshifts in 
the models of interest, it is possible that the unstable fluctuations do not have time to grow significantly.  We defer this question to future work.

\subsection{Flat space dynamics}

Before coupling the theory to gravity, it is useful to understand its dynamics in flat space.  
 The canonical momentum is 
\be
    \pi = {df\over d\dot\phi} = f'\dot\phi\,,
\ee
hence the Hamiltonian is
\be
    {\cal H} = \pi\dot\phi - {\cal L} = 2 f'X - f + V.
\ee
Consider the example of quadratic $f$,
\be
    f(X) = \sfrac12\alpha(X-X_0)^2
    \label{quadf}
\ee
where $\alpha$ and $X_0$ are constants.  The conserved energy is then
\be
    {\cal H} = \sfrac12 \alpha (X-X_0)(3X+ X_0) +  V
    \label{conservedH}
\ee
The system can be reduced to quadrature by solving for $\dot\phi$, 
\be
    \dot\phi = \pm\left(\frac23\left(X_0\pm 2
    \sqrt{X_0^2 +{3\over 2\alpha}({\cal H}-V)}\right)\right)^{1/2}\,.
    \label{dotphieq}
\ee
One can then numerically integrate $dt = d\phi/\dot\phi$ to obtain $t(\phi)$.
For large $({\cal H}-V)$ there are four branches, but at sufficiently small energies, the argument of the square root becomes negative and there are only two.  

\begin{figure}[t]
\centerline{\includegraphics[width=\columnwidth]{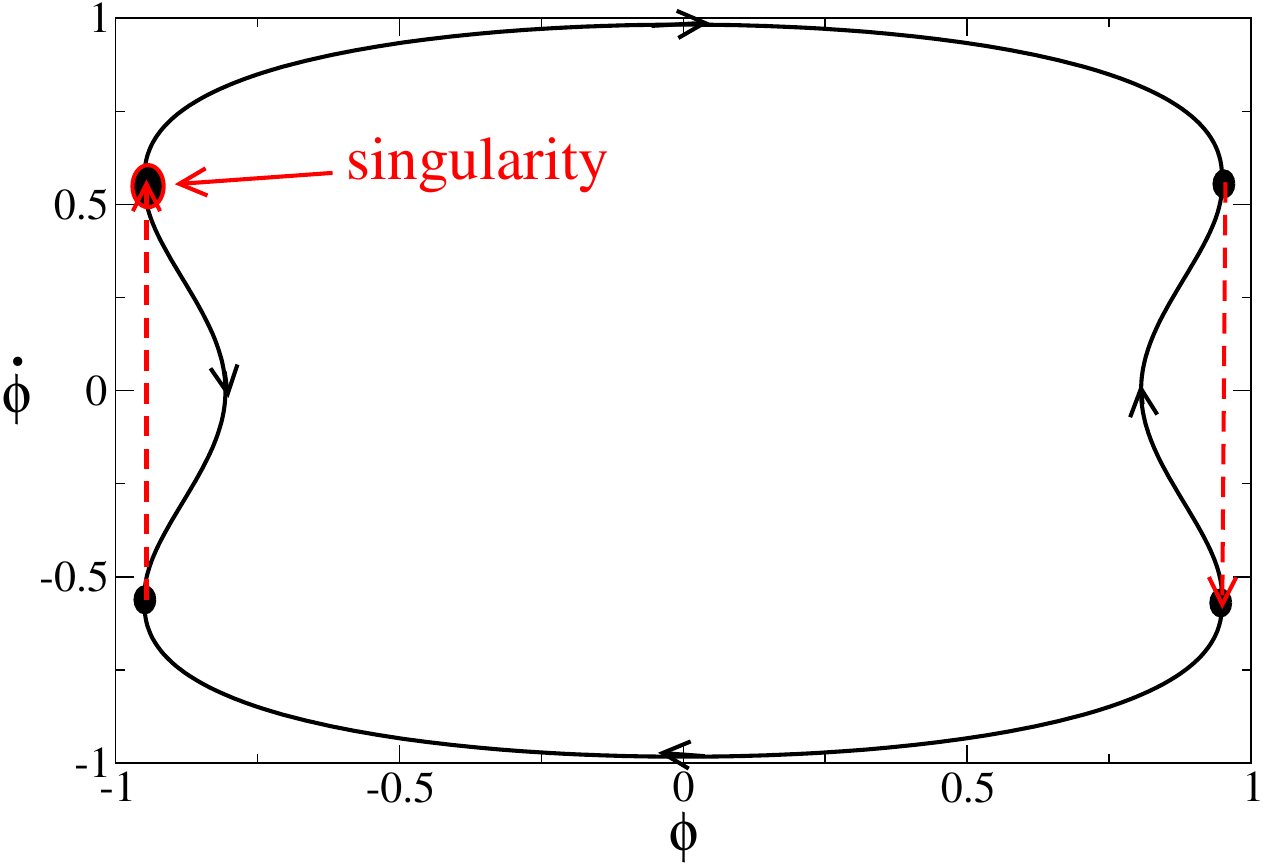}}
\caption{Phase space trajectories for the quadratic $f(X)$ model (\ref{quadf}). Arrows denote the direction of time, and red dashed lines indicate the momentum discontinuity when the
system hits the ``brick wall''.  Point labeled ``singularity'' denotes the beginning of the cosmological solution discussed below.
} 
\label{fig:ppplot}
\end{figure}

\begin{figure*}[t]
\centerline{\includegraphics[width=\columnwidth]{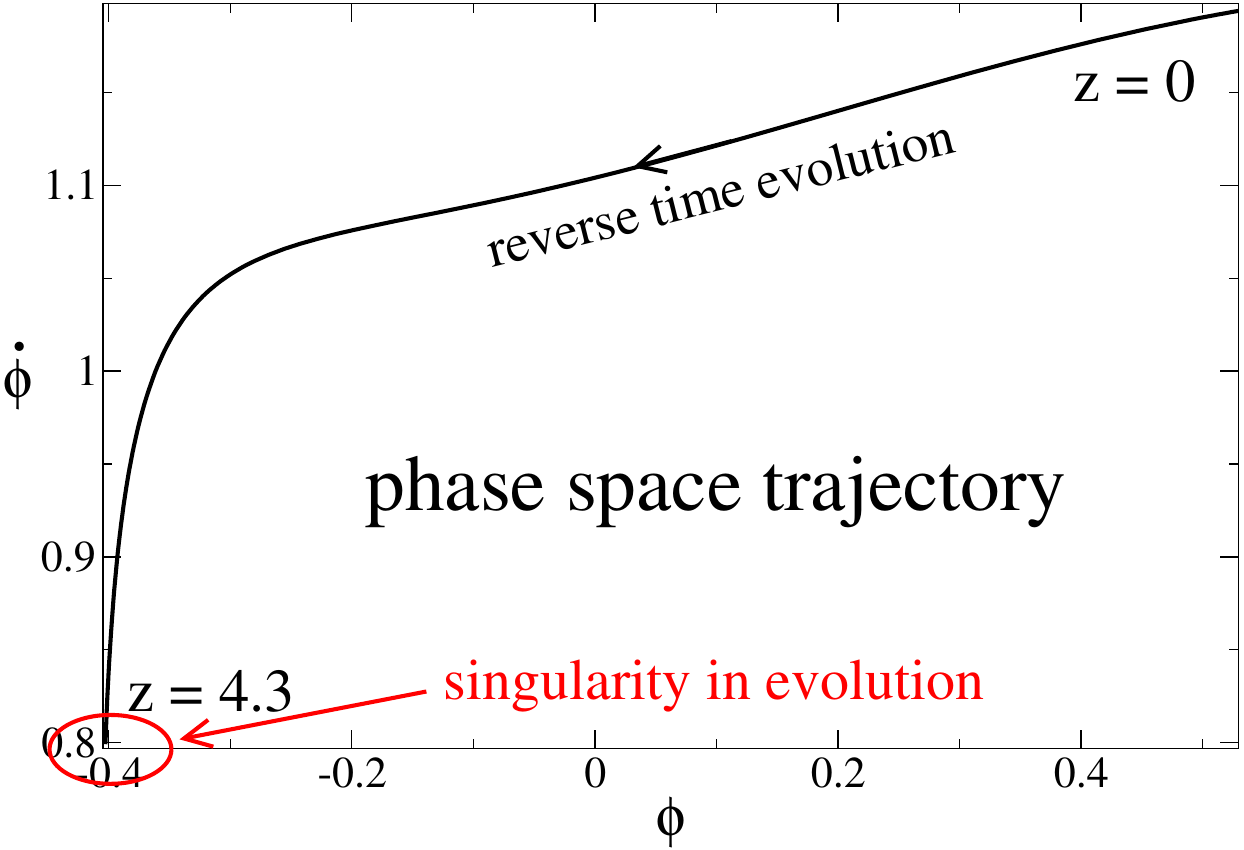}\includegraphics[width=1.05\columnwidth]{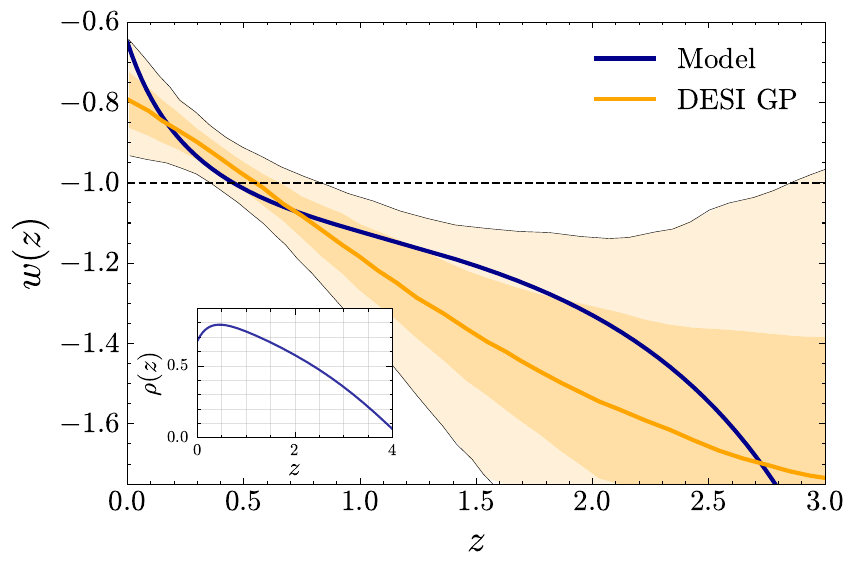}}
\caption{Left: phase space evolution of solution for 
quadratic $f(X)$ model, indicating reverse time evolution from the present to $z=4.3$.
Right: $w(z)$ for the MCMC best-fit solution (blue) for the same model, compared to DESI best-fit GP determination \cite{DESI:2025fii} (orange). The shaded regions are the DESI 1$\sigma$ and 2$\sigma$ confidence intervals. 
Inset shows the evolution of the dark energy with $z$.
} 
\label{fig:wquadplot}
\end{figure*}

The phase space diagram for the case of four branches is illustrated in Fig.\ \ref{fig:ppplot};
the case of two branches corresponds to removal of the curves that connect negative and positive values of the velocity at the heavy dots.\footnote{In Fig.\ \ref{fig:ppplot} we have assumed an unstable potential $V = -\sfrac12|m^2|\phi^2$, motivated by the cosmological data, as will be described below.}
Arrows 
denote the flow of time.  The time evolution leads to the system moving toward the dots in the first or third quadrants.  At these points, there is no possibility for further evolution, unless the system suddenly reverses velocity and jumps to the corresponding dot in the fourth or second quadrant.
This behavior was noted in Ref.\ \cite{Shapere:2012nq}, which proposed such models in the context of condensed matter systems and dubbed them ``time crystals''.    The sudden velocity reversal was likened to a ball bouncing off a brick wall.  Such a discontinuous change might be considered an approximation to continuous evolution in some more UV-sensible theory.  It will turn out that we do not have this behavior in the cosmological solutions discussed below: the system must emerge from either the quadrant-two or -four dot at some redshift $z\sim 4$, with no explanation for its dynamics at earlier times.  

It is easy to see that phantom crossing can occur in such a model.  The equation of state is given by
\bea
    w &=& {p\over\rho} = {{\cal L}\over{\cal H}} = -1 + 2 X {f'\over {\cal H}}\,.
     \label{wgenrel}
\eea
We used Eq.\ (\ref{conservedH}) to eliminate $V$ from the numerator to obtain the final form of (\ref{wgenrel}).  Since
$X>0$, it is clear that phantom crossing occurs whenever
$f'$ changes sign.  For the quadratic $f$ model,
this means $X$ crossing below $X_0$.  Moreover, by
letting ${\cal H}\to 0$, $w$ can be made arbitrarily negative.  The general result (\ref{wgenrel}) shows why the linear phantom model with $f' = -1$ can never cross $w=-1$, except by having a pole in $w$ where ${\cal H}$ changes sign.

\subsection{Cosmological dynamics}
\label{cosmosect}
Putting the model into a Friedmann-Lema\^\i tre-Robertson-Walker (FLRW) cosmological background with scale factor $a(t)$, the classical equation of motion (EOM) is
\be
    \ddot\phi\left(f' + 2X f''\right) + 3Hf'\dot\phi = -V'\,,
    \label{eom}
\ee
where $H=\dot a/a$ is the Hubble parameter.
For numerical evolution, it is more convenient to change the independent variable from cosmological time $t$ to $a$, and to rewrite the EOM in Hamiltonian form,
\bea
    \label{foeom}
    \pi' \left(f' + 2X f''\right)  + 3f' \pi &=& -V'/aH\\
    \phi' &=& \pi/aH\nn\,,
\eea
where primes on $\pi$ and $\phi$ denote $d/da$, and the Hubble parameter is defined by 
\bea
    H^2 &=& \sfrac13 8\pi G(\rho_m + {\cal H})\nn\\
       &=& H_0^2\left(\Omega_m a^{-3} + \rho_c^{-1}{\cal H}\right)
\eea
in terms of the critical density $\rho_c$.  We have ignored the radiation density $\rho_r$ since we are interested in the relatively late Universe.  In the following we will adopt geometrized units $8\pi G = 3$, and take $\rho_c = 1$ as the unit of energy density, which also implies $H_0=1$.  Using the freedom to rescale $\phi\to Z\phi$ with arbitrary $Z$, we can set $X_0=1$ in these units, which eliminates it as a free parameter.

As benchmarks for explaining the DESI observations, we aim to find models giving rise to expansion histories that are close to the best-fit reconstructed $w(z)$ curve obtained in Ref.\ \cite{DESI:2025fii}, which used the Gaussian process (GP) algorithm.  This is most straightforward to do by imposing initial conditions at the present time, $a=1$, and evolving the system backwards in time.  The DESI best-fit model has $\Omega_m \simeq 0.31$, and so we take ${\cal H}_0 = 1 - \Omega_m = 0.69$.  Moreover, it has
$w_0 = -0.75$ and slope $w_0' = 0.5$.  We can then solve Eq.\ (\ref{wgenrel}) for the initial value $\pi_0$ of the field velocity.
To determine the initial field value $\phi_0$, one can use the general relation
\be
    V = f - {w\over 1+w} 2 X f'\,,
    \label{Veq}
\ee
which is equivalent to Eq.\ (\ref{wgenrel}). By evaluating it at $a=1$, $\phi_0$ can be determined.

We would also like to match the slope $w_0'$ at $z=0$, within errors.
Differentiating Eq.\ (\ref{wgenrel}) to derive a relation between $w'_0$ and the derivative of the potential,
\bea
    V' = \frac{-\pi Hf'}{2(f' + X f'')} \Bigg[
        &\, 3\Big( f'(1 - w)- 2 X w f'')\Big) \nonumber \\
        & + \frac{w'}{1 + w} (f' + 2 X f'') 
    \Bigg]\,.
    \label{Vpeq}
\eea
By doing this matching at $a=1$, we can solve for one of the parameters in the potential.  This procedure is useful for searching by trial and error for models that provide a good fit to the data.  These can serve as a starting point for a more systematic Monte Carlo scan over parameters, as we will describe in Section \ref{mcmc}.

A simple choice for the potential is
\be
    V = \Lambda + \sfrac12 m^2\phi^2\,.
\ee
Carrying out the procedure described above, we find that $m^2$ is always negative when matching $w'_0$,
corresponding to a potential that is unbounded from below.  This tachyonic instability is a relatively benign pathology compared to all the rest of the model's peculiarities, which in any case could be fixed by adding a $\lambda\phi^4$ term.  (We will mention more general choices of $V$ later on.)
The field starts on one side of the hilltop and crosses to the other between $z\sim 4$ and the present time.  The phase space plot is shown in
Fig.\ \ref{fig:wquadplot} (left).

The remaining free parameters of the model are $\alpha$ and $\Lambda$.  By trial and error, we find examples that lie mostly within the 1-$\sigma$ error band of the DESI GP curve, as illustrated in 
Fig.\ \ref{fig:wquadplot} (right).  A generic feature of the solutions is that a singular point in the reverse time evolution is reached at a finite redshift, which cannot be made much larger than $z\sim 4$, even with careful tuning of the parameters.  

The endpoint can be understood from
Fig.\ \ref{fig:ppplot}: it corresponds to the circled point in the second quadrant.  If energy was conserved, the evolution could be continued by reversing the field
velocity and continuing to integrate along one of the two possible branches.  However in the expanding universe, the dark energy is not conserved, and in fact it is decreasing toward the past since $w < -1$.  At the singular endpoint, the argument of the square root in Eq.\ (\ref{dotphieq}) has reached zero.  In a static universe, sudden reversal of $\dot\phi$ would enable it to start increasing again so that the evolution could continue.  But in the cosmological
setting, ${\cal H}-V$ is continuing to decrease at early times, so that the argument of the square root
 becomes negative, and no further evolution is possible.\footnote{It is straightforward to show that $d{\cal H}/du = -6Xf'$ for the cosmological
 solutions, which decreases in the past for $f'<0$, corresponding to $w<-1$ as expected.}

This leads to the peculiar feature that the ghost condensate model solutions emerge from nothing at
some finite redshift, which at best can be made larger than the values to which current data are sensitive.  One might hope that some ultraviolet completion would enable the solutions to extend back to earlier times.  Here we take a phenomenological viewpoint, based on the fact that the dark energy density is starting out at negligible values at the time of its singular emergence.  We can therefore reasonably ignore its effects at earlier times.  
The solutions that we present take the upper trajectory leading away from the singularity in Fig.\ \ref{fig:ppplot}.   An animation of the time-reversed evolution is available online \cite{https://doi.org/10.5281/zenodo.16728376}.

Despite having several free parameters ($\alpha$, $\Lambda$ and $m^2$), it is not possible to closely match the shape of the central DESI $w(z)$ curve.
At best, the solution undulates around this central curve, while trying to stay within the 1-$\sigma$ preferred region.  At large $z$, it always diverges,
either due to ${\cal H}$ crossing zero, or as a result of the singularity (which may happen slightly earlier than ${\cal H}$ crossing zero).

The aforementioned features in $w(z)$ appear to be a robust prediction of this class of models.
We have investigated alternative functions $f(X)$,
including $\alpha X(X-X_0)^2$, $\alpha (X-X_0)^p$, and $\sfrac12\alpha(X-X_0)(X-\beta X_0)$.  These typically give no better results for $w(z)$, or only a marginal improvement.  The same can be said for generalized potentials $V = \Lambda +\sfrac12 m^2\phi^2 + g\phi^3 + \lambda\phi^4$.  The small improvement in the fit to cosmological data is insufficient to justify the addition of new parameters.

\section{Coupled DM-DE models}
A less exotic possibility for apparent phantom divide crossing is to let the dark matter (DM) mass 
be a function of $\phi$, which could occur through renormalizable operators such as
$\phi^2 S^2$ for scalar DM or $\bar\chi\phi\chi$ for
fermionic DM \cite{Farrar:2003uw}.  Such operators become time-dependent contributions to the quintessence potential when we relate $S^2$ or $\bar\chi\chi$ to the dark matter density.
The resulting shift in the dark energy potential takes the form
\be
    \Delta\rho = {\rho_{dm,0}\over a^3}\left({g(\phi)}-1\right)\,,
\ee
where $\rho_{dm,0}$ is the present-day dark matter density, and $\rho_{dm,0}\,g(\phi)$ is its
value at earlier times.  The function $g(\phi)$ can be expressed as\footnote{The potential energy density of scalar dark matter $\sfrac12m^2\phi^2$ is quadratic in its mass, while that for fermionic
DM $m \bar\chi\chi$ is linear.}
\be
    {g(\phi)} = \left\{
    \begin{array}{ll} M^2(\phi)/M_0^2, & \hbox{scalar DM}\\
        M(\phi)/ M_0, & \hbox{fermionic DM}
        \end{array}\right.\,,
\ee
where $M(\phi)$ is the DM mass and $M_0$ is its present value.  The extra contribution $\Delta\rho$ contributes to the apparent dark energy density, when matter is assumed to have a constant mass, but it does not contribute to the pressure since cold dark matter is pressureless.  Therefore the apparent dark energy equation of state is
\be
    w_{\rm eff} = {\sfrac12\dot\phi^2 - V\over
        \sfrac12\dot\phi^2 + V +  \Delta\rho}
        = {w_\phi\over 1+ \Delta\rho/\rho_\phi}\,,
\ee
where $w_\phi$ and $\rho_\phi$ are the usual expressions from the case of no interaction with DM.  If $\Delta\rho/\rho_\phi<0$ in the past and $w_\phi < 0$, it is possible to make $w_{\rm eff} < -1$. {If one wishes to match $w_0$ and $w_0'$ today, Eq.\ (\ref{Veq}) remains valid in this class of models,
with $f=X$ and $f'=1$, while Eq.\ (\ref{Vpeq}) becomes
\be
V_0' + \left(1-\sfrac12 w_0\right)\Delta\rho'_0 = 
-{\pi_0\over 2}\left(w_0' + 3(1-w_0^2)\over (1+w_0)\right)\,.
\ee}

\begin{figure*}[t]
\centerline{\includegraphics[width=\columnwidth]{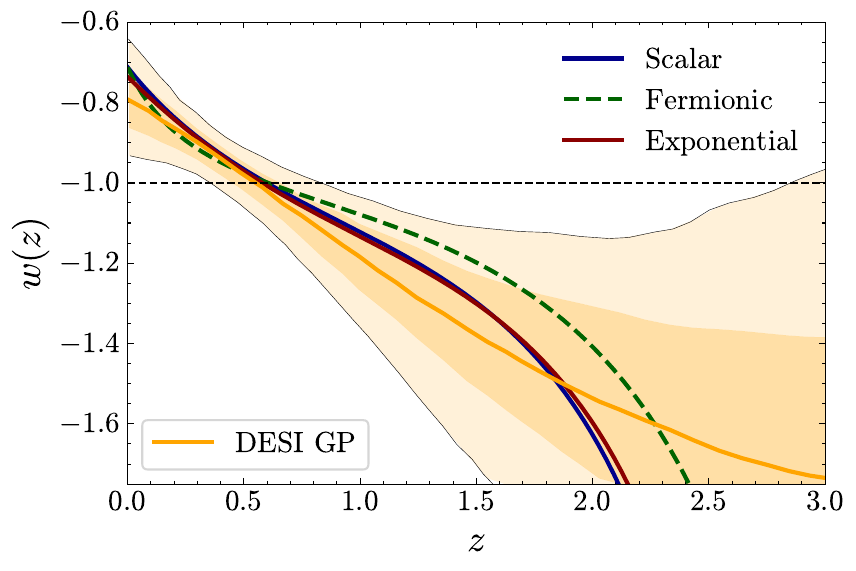}\includegraphics[width=0.97\columnwidth]{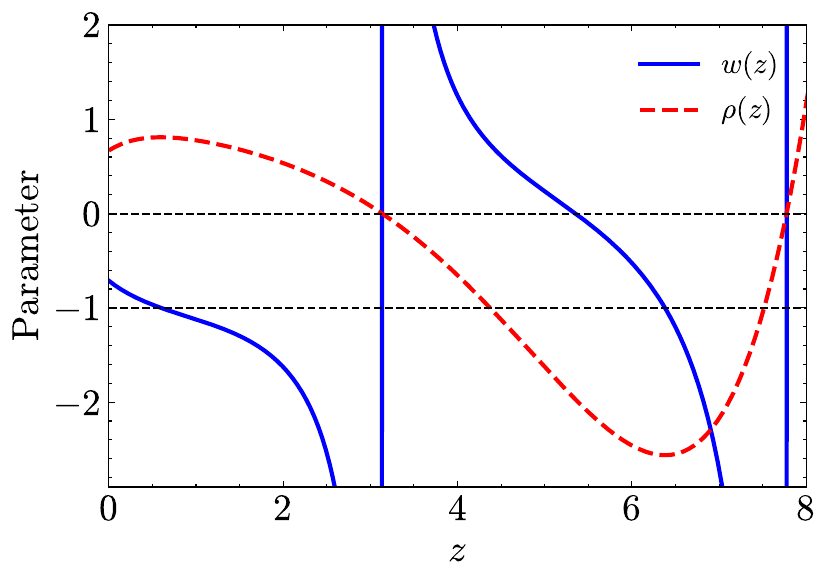}}
\caption{Left: Best-fit equation of state ($w$) versus redshift ($z$) for the three DM-DE interaction models, compared to the DESI Gaussian Process (GP) curve. The shaded regions show the DESI 1$\sigma$ and 2$\sigma$ intervals. 
Right: Equation of state and dark energy density ($\rho$) as a function of redshift for the coupled DM-DE model with scalar potential (\ref{masspot}). The EOS diverges when the effective dark energy density vanishes, twice.} 
\label{fig:wdmde}
\end{figure*}

To test these models against the DESI data, we use the same strategy as described in Section \ref{cosmosect}, to initially search by hand for promising solutions. Boundary conditions at $a=1$ are determined by demanding that $\rho_\phi = 1-\Omega_m \cong 0.69$ and $w_0 \cong -0.75$, while one model parameter is
constrained by setting $w_0'\cong 0.5$.  The equations of motion are then integrated backwards in time.  Carrying this out for the simple case where
\bea
V &=& \Lambda +\sfrac12 m^2\phi^2\nn\\
\Delta\rho &=& \sfrac12\mu^2(\phi^2-\phi_0^2)/a^3\,,
\label{masspot}
\eea
we obtain an optimal $w(z)$ curve that is surprisingly similar to the one found in the quadratic $f(X)$ models.  Unlike those models,
the potential is stable in this case, with $m^2\cong 3.2$, while $\mu^2 \cong -0.25$ for
$\Lambda = 0.17$.  

In the case of fermionic dark matter, 
we replace Eq.\ (\ref{masspot}) by
$\Delta\rho = A(\phi-\phi_0)/a^3$ \cite{Costa:2014pba}.  A good fit to the data is found with
$m^2=-3.79$, $A = 0.10$, $\Lambda = 1.2$, and initial conditions $\phi_0 = 0.561$, $\dot\phi_0 = 0.415$.  Like most of the models we studied in the DM-DE interaction class,
the dark energy crosses zero twice at higher redshifts, leading to poles in $w(z)$ near $z=4$ and $z=9$.

In the DM-DE interaction models, there is no singularity in the time-reversed evolution; instead poles appear in $w(z)$ at redshifts where $\rho_\phi + \Delta\rho$ crosses zero.  The same kind of tuning of parameters as needed in the $f(X)$ models must be done here to push these poles far enough into the past to not spoil the desired $w(z)$ shape.  This behavior is generic because $w_{\rm eff}<-1$ drives $\rho_\phi + \Delta\rho$ to 
smaller values in the past, and in general there is no reason for it to not fall below zero.  However we find that it starts rising again at earlier times so that there is a second pole, and $w_{\rm eff}$ tends toward a constant positive value in the far past.

A somewhat better fit can be obtained using exponential potentials,\footnote{Similar models were considered in Refs.\ \cite{Wetterich:1994bg,Amendola:1999er,Pettorino:2013oxa,2012PhRvD..86j3507P}}
\bea \label{eq:exp_potential}
    V &=& \Lambda + A e^{-k\phi}\nn\\
    \Delta\rho &=& B a^{-3}(e^{-k\phi}-1)\,.
\eea
In this model we have the freedom to define $\phi_0 = 0$ today since different choices amount to a rescaling of $A$ and $B$.  An example that falls inside the DESI 1-$\sigma$ region has $A = 0.60$,
$B=-0.028$, $\Lambda = 0.005$, and $k=2.64$.  Like in the model (\ref{masspot}), $\Delta\rho$ crosses zero twice, leading to poles in $w(z)$ 
near $z=4.5$ and $12.5$.

\section{Likelihood Analysis} \label{mcmc}
Having identified the qualitative features of the models of interest that are likely to give good fits to the DESI data, we compute their likelihoods. We use the Boltzmann code \texttt{CAMB} \cite{Lewis:1999bs} to study the modified evolution history and use the parametrized post-Friedmann approach \cite{PhysRevD.78.087303} to calculate dark energy perturbations. We then employ Markov Chain Monte Carlo sampler \texttt{Cobaya} \cite{Torrado:2020dgo} to constrain the free parameters and compute the likelihoods. 

\begin{figure*}
\centerline{\includegraphics[width=\columnwidth]{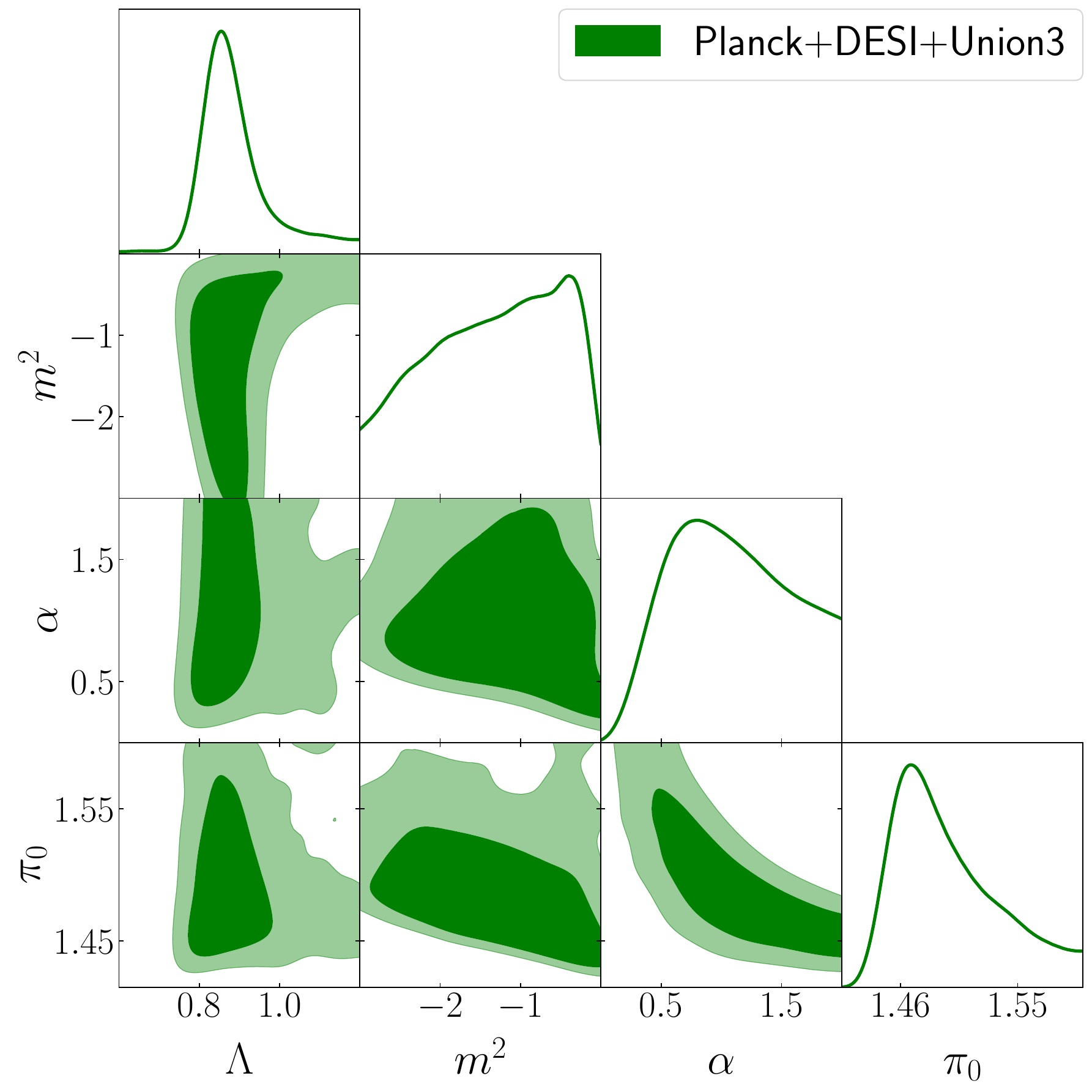}
\includegraphics[width=\columnwidth]{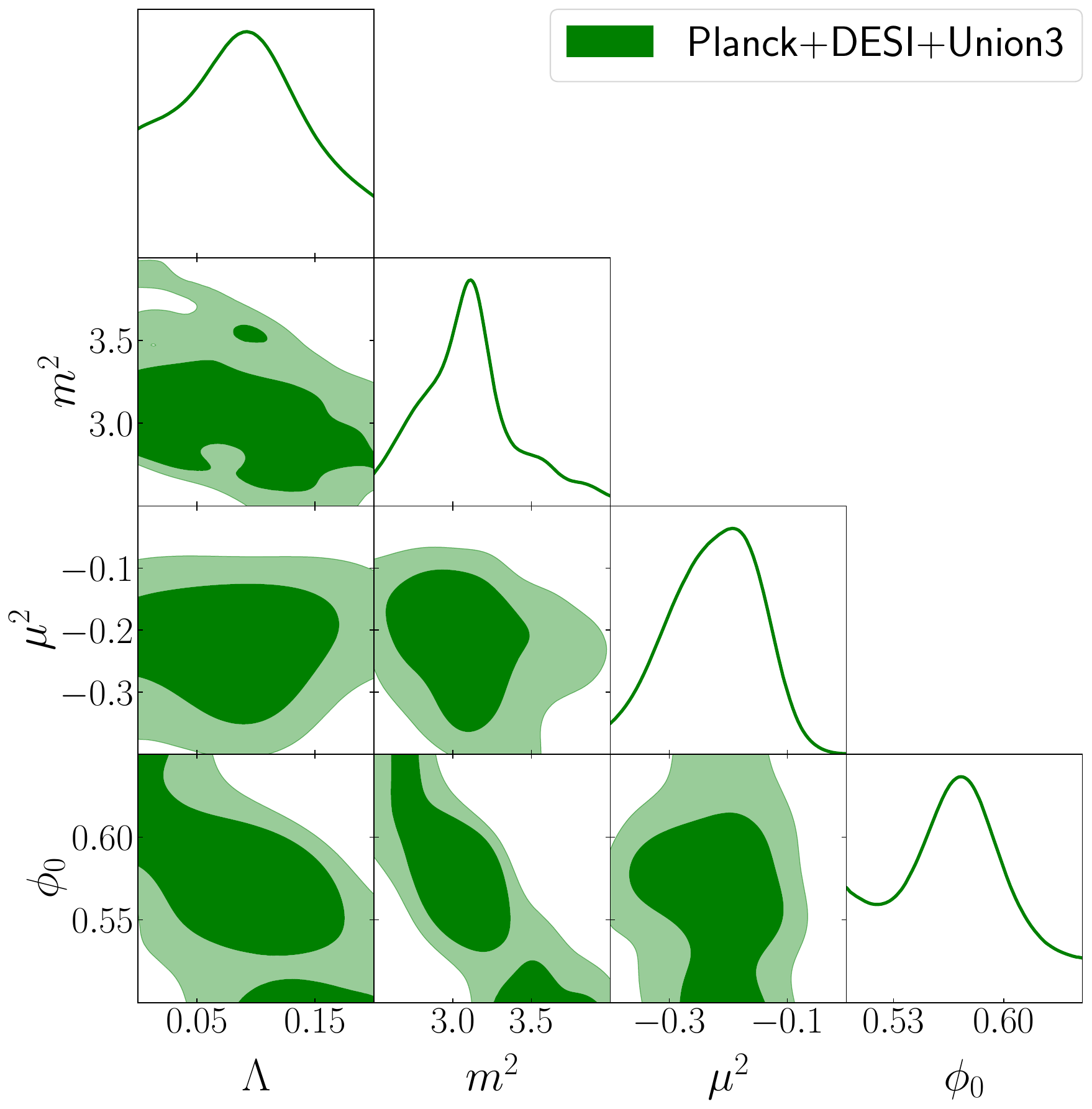}}
\caption{Marginalized Posteriors on parameters for the quadratic GC (left) and  and Scalar DM-DE models (right).  The free parameters are  $V_0=\Lambda, m^2, \alpha, \pi_0$ for GC and  
 $\Lambda, m^2, \mu^2, \phi_0$ for scalar DM-DE.} 
\label{fig:triangle12}
\end{figure*}
\begin{figure*}
\centerline{\includegraphics[width=\columnwidth]{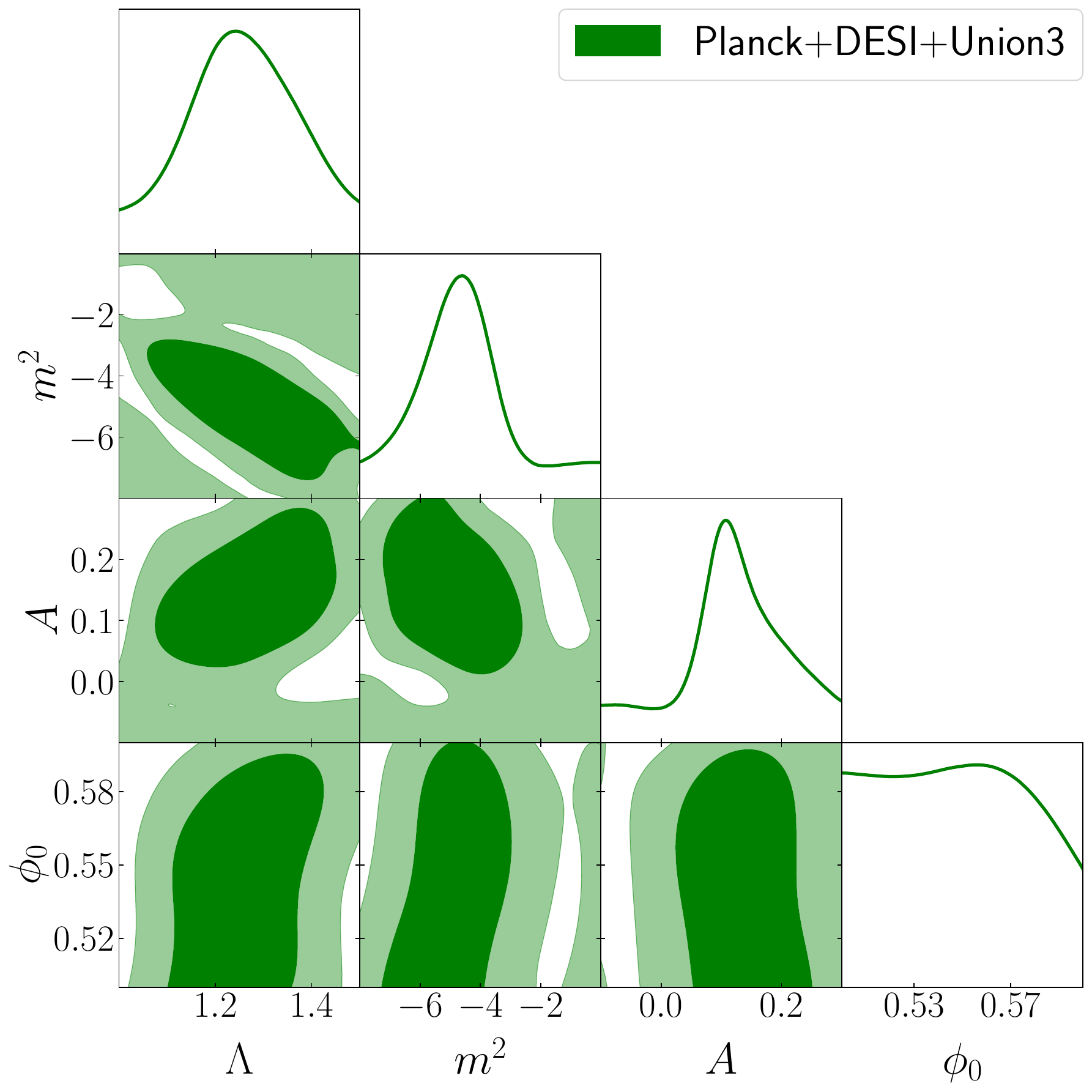}
\includegraphics[width=\columnwidth]{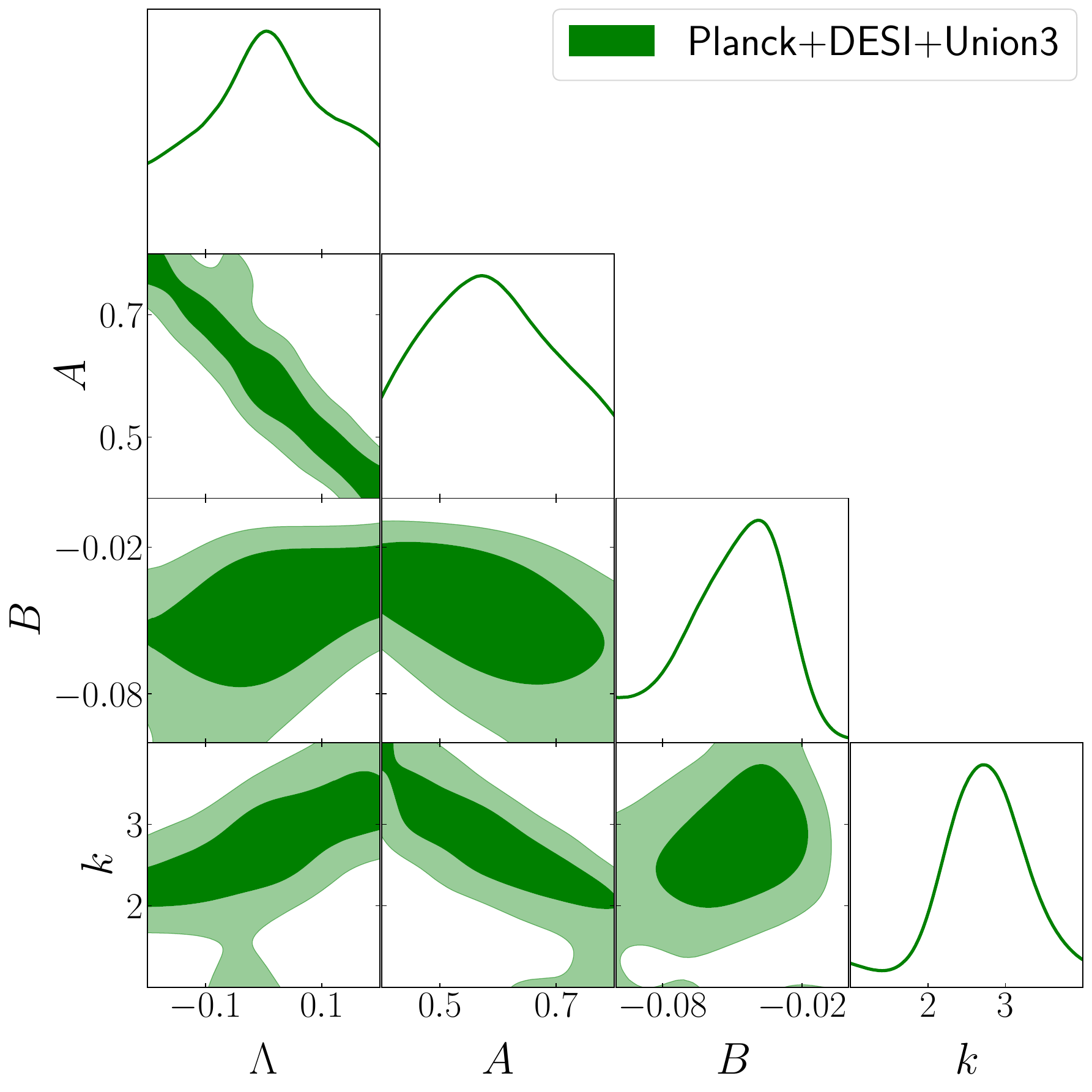}}
\caption{Marginalized Posterior on parameters of the Fermionic (left) and Exponential (right) DM-DE models, which are respectively $\{\Lambda, m^2, A, \phi_0\}$ and $\{\Lambda, A, B, k\}$.} 
\label{fig:triangle34}
\end{figure*}

We adopt the datasets employed in the DESI paper \cite{DESI:2025fii}, namely Planck 2018 anisotropies \cite{Planck:2018vyg} and lensing \cite{Planck:2018lbu}, DESI 2025 BAO \cite{DESI:2025zgx}, and Union3 SN \cite{Rubin:2023jdq}. Convergence of the chains was monitored using the Gelman-Rubin statistic \cite{Gelman:1992zz}, demanding $R-1 \leq 0.02$. For the Ghost Condensate (GC) model with quadratic $f(X)$, the field was evolved backwards in time up to the singularity,  beyond which we consider $w_\phi<-1$ to be constant and $\rho_\phi \sim a^{-3(1+w)}$, hence continuing to decrease toward the past. 

For the DM-DE interaction models, $w_\phi$ starts to increase again (toward the past) after the initial period of phantom crossing.  This is inevitable: $\rho_\phi$ must decrease toward the past while $w_\phi<-1$, but its potential is bounded from below, so eventually it must start increasing again.  This could either happen after $\rho_\phi$ has become negative, 
leading to two zero-crossings and hence two poles in $w_\phi(z)$, or before then, avoiding any poles.  But in either case $w_\phi$ eventually increases and crosses zero,
typically around $z\sim 3.5-4$ for models that provide a good fit to the DESI data.  For these models, we therefore stop the time-reversed evolution when $w_\phi=0$. This is a safe choice from the standpoint of deriving constraints, since by this time the dark energy has already reached a negligible level and has no effect on observables. 

\begin{center}
\renewcommand{\arraystretch}{1.4}
\begin{table*}
\begin{tabular}{|c|c|c|c|c|l|l|l|c|}
\hline
Model & $\Delta \chi^2 $ & $H_0$ & $\Omega_m$ & $\Lambda$ &\multicolumn{3}{|c|}{Other parameters} & $\beta(z)$\\
\hline
GC& $-15.2$& $66.17^{+0.83}_{-0.98}$ & 0.325 & 0.83 & $m^2: -1.53$ & $\alpha: \phantom{+}1.10$ & $\pi_0: \phantom{+}1.482$ & $-$ \\
\hline
Scalar & $-12.1$ & $66.55^{+0.81}_{-0.81}$ & 0.330 & 0.09 & $m^2: \phantom{+}3.09$ & $\mu^2: -0.13$ & $\phi_0: \phantom{+}0.559$  & $\gtrsim 0.6$\\
\hline
Fermionic & $-13.3$ & $66.62^{+0.70}_{-0.93}$ & 0.323 & 1.18 & $m^2: -3.76$ & $A: \phantom{+}0.016$ & $\phi_0: \phantom{+}0.565$ & 0.03\\
\hline
Exponential & $-15.1$ & $66.75^{+0.96}_{-0.96}$ & 0.328 & $-0.02$& $k: \phantom{++}2.61$ & $A: \phantom{+}0.60$ & $B: -0.027$  & $0.006-0.01$ \\
 \hline  
\end{tabular}
\caption{$\Delta\chi^2$ relative to $\Lambda$CDM of the Ghost Condensate (GC) model with quadratic $f(X)$ and the three DM-DE interaction models, the $1\sigma$ range of the Hubble parameter ($H_0$) in km s$^{-1}\text{Mpc}^{-1}$, and best-fit values of $\Omega_m, \Lambda$ and other free parameters.  The last column shows the ratio $\beta(z)$ for $z\in[0,10]$ of the fifth force between DM particles mediated by quintessence, relative to gravitational strength, where applicable.} 
\label{table:table1}
\end{table*}
\end{center}

Henceforth, we refer to the ghost condensate model and the three DM-DE potential cases as GC, Scalar, Fermionic, and Exponential, respectively. All models have four additional free parameters compared to $\Lambda$CDM.  These include the initial field velocity ($\pi_0$) for GC and the initial field value ($\phi_0$) for the Scalar and Fermionic models,  in addition to the Lagrangian parameters. Unlike the previous sections where the initial field velocity  $\pi_0$ was determined in terms of presumed values for $w_0$ (Eq.\ (\ref{wgenrel})), here we fix $\pi_0$ by demanding that
${\cal H}_0 = 1 - \Omega_m$ from imposing spatial flatness, and treating $\phi_0$ as a model parameter.\footnote{With the exception of the Exponential DM-DE model, where $\phi_0\equiv 0$ by appropriate rescaling of $A$ and $B$} We assume a uniform prior on the parameters centered around the values mentioned in the previous section, which were determined through trial and error. 

In Table \ref{table:table1}, we list the improvement in $\chi^2$ relative to $\Lambda$CDM
for each of the four models.  The Ghost Condensate (GC) and Exponential DM-DE models
achieve $\Delta\chi^2 = -15$, which is essentially equal to the result found by DESI
\cite{DESI:2025fii} for their best-fit Gaussian Process reconstruction.  The other two models perform only slightly worse.  

We do not try to assign confidence levels to the preference over $\Lambda$CDM since the models have four additional parameters.  It is not clear that they should be penalized by this number, using an information criterion, in the same way as one would do for an ansatz for phenomenologically fitting the equation of state.  One reason is that there are degeneracies between model parameters, as can be seen from the posterior distributions in 
Figs.\ \ref{fig:triangle12} and \ref{fig:triangle34};
therefore, it is not necessary to have all four of them varying freely to find a good fit to the data.  Conversely, we have experimented with models having even more free parameters, for example by adding extra terms to the $V(\phi)$ potential or different functions $f(X)$ in the GC models, and we find that they do not give any significant improvement.  The best-fit values for the various parameters are given in  Table \ref{table:table1}.  All of the models prefer a value of Hubble parameter near $66-67 ~\text{km s}^{-1}\text{Mpc}^{-1}$, which pulls it farther away from the local measurement \cite{Riess:2020fzl}.

\section{Dark matter long-range forces}

The coupling of dark matter to an ultralight quintessence field can introduce a long-range force between DM particles, that is constrained  by its effect on structure formation \cite{1992ApJ...398..407G}.  For example, fermionic DM with a coupling $y\phi\bar\chi\chi = y\phi n_\chi$ corresponds to $A\phi$ in the parametrization we used, hence $y = m_\chi A/\rho_\chi$.  It gives rise to a long-range potential
between DM particles that can compete with gravity. To quantify the strength of such interactions, we need to convert the geometrized units to physical ones.  $A$ has units of $\rho_c$ divided by scalar field,
and the units of $\phi$ are 
\be
   [\phi]\equiv \left(8\pi G\over 3\hbar c^5\right)^{-1/2} = 4.2\times10^{18}\,{\rm GeV}\,,
\ee
since we have been taking $8\pi G/3 = 1$.  (Hence $[\phi]$ is $\sqrt{3}$ times the reduced Planck mass.) Therefore
\be
    y = {m_\chi A \rho_c\over [\phi]\rho_\chi} = \left(A\over [\phi]\Omega_c\right)  m_\chi = {m_\chi\over 7\times 10^{19}\,{\rm GeV}}\,.
\ee
If $\phi$ is approximately massless, then the potential induced by its exchange between two DM particles is $V = -y^2/(4\pi r)$.   The strength of this interaction relative to gravity,
which we will denote by $\beta$, is
\be
   \beta =  {y^2 \over 4\pi G\, m_\chi^2} = 0.002\,,
\ee
which is allowed by the constraint $\beta < 0.03$ found in Ref.\ \cite{Archidiacono:2022iuu}.  Since the coupling is independent of $\phi$, this ratio is constant while the field evolves.

\begin{figure*}
\centerline{\includegraphics[width=2\columnwidth]{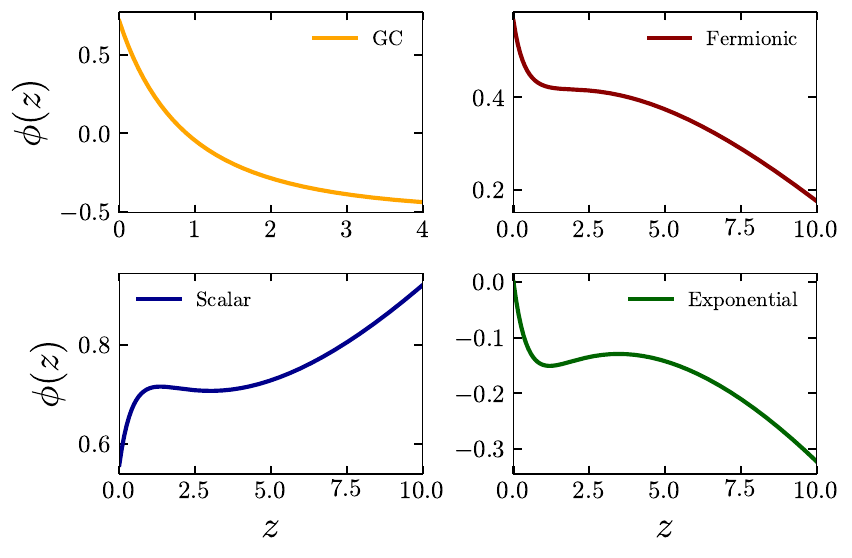}}
\caption{Evolution of the quintessence field with redshift, for the best-fit parameters in each of the four models considered.} 
\label{fig:solns}
\end{figure*}

For scalar DM, consider the microphysical coupling $\sfrac12\lambda S^2\phi^2 = \sfrac12\mu^2 \phi^2$.  In this case, we identify $m_s^2 S^2$ with $\rho_s$, the DM energy density.  Then following similar reasoning as above, the relative strength of the potentials from the $\phi$ mediated force and gravity is 
\bea
   \beta(z) &=& \left(\lambda \phi(z)\over m_S^2\right)^2{1\over 4\pi G}\nn\\ &=&  \left(\mu^2\phi(z)\over \Omega_c[\phi]^2\right)^2{1\over 4\pi G} \ge 0.05
   \label{scalar-beta}
\eea
which is excluded.  Unlike the fermionic case above, it varies with redshift because the linearized coupling of $\phi$ to scalar DM depends upon $\phi$.  
In the above expressions, $A$, $\mu^2$ and $\phi(z)$ keep their dimensionless values and $[\phi]$ supplies the explicit factors of energy.   Fig.\ \ref{fig:solns} show that $\phi$ was larger in the past for this model, hence the inequality in (\ref{scalar-beta}), which is saturated at $z=0$.

Similarly, for the Exponential DM-DE model, we assume a microscopic interaction
\be
    {\cal L } = \lambda e^{-k\phi}\left\{\begin{array}{ll} m_\chi\bar\chi\chi, &\hbox {fermionic DM}\\
    m_S^2 S^2, & \hbox{bosonic DM}\end{array} \right.\,.
\ee
With this choice, $\lambda = B/\rho_{\sss DM} \approx 0.1$ for either kind of DM candidate. Restoring units, we find
\be
    \beta(z) = \left(B k e^{-k\phi(z)}\over \Omega_c[\phi]\right)^2{1\over G} \sim (5-8)\cdot10^{-4}\,,
\ee   
where the range of values between $z=0$ and $z=10$ 
relevant for structure formation is shown.
Therefore this kind of model is safe from the fifth force constraint for either bosonic or fermionic dark matter.

\section{Conclusions}

In this work we studied two classes of single-field quintessence models that are capable of crossing the phantom divide.  Even though the models look very different from each other, they both predict qualitatively similar histories for the dark energy equation of state $w_\phi(z)$.
Interestingly, the curvature of $w(z)$ predicted by the models (convex) is opposite to that favored by the DESI
analysis \cite{DESI:2025fii}, which is concave.  
Nevertheless,  they still provide significant improvement over $\Lambda$CDM, since the confidence regions around the DESI best-fit Gaussian Process  are still relatively wide.  Perhaps future data will be able to distinguish between the convex and concave shapes.

We point out that the convex shape for $w(z)$ is a natural expectation for models with $w_\phi<-1$ at early times,
from the perspective of evolving backwards in time.
$\rho_\phi$ must decrease in this direction, and unless it is forbidden from crossing zero, a pole in $w(z)$ will appear, which has the convex shape.  Even if $\rho_\phi$ does not cross zero, $w(z)$ tends to have the same shape.
Hence if the concave shape should be favored by future data, it will present even greater model-building challenges.

The concept of coupling quintessence to dark matter seems simple and natural from the theoretical perspective, and it may be the most promising approach if the experimental indications for crossing below $w=-1$ persist.  It is encouraging that the simplest expectations for the couplings, renormalizable interactions $\phi^2 S^2$ for scalar DM $S$ and $\phi\bar\chi \chi$ for fermionic DM $\chi$, give good fits to the data; however the former is ruled out by constraints on long range forces between DM particles.  
On the other hand,
exponential couplings of $\phi$ to DM lead to a sufficiently weak force as to still be allowed by 
current studies of structure formation.  These assertions should be checked by future studies of structure formation that explicitly take into account the evolution of the coupling strength of $\phi$ to DM in the cases where it is not constant.
It is intriguing that the nature of the DM particle could in this way be probed by the expansion history of the Universe, even if it has no interactions with the particles of the Standard Model.

A further remark in favor of the Exponential DM-DE interaction model can be made.  According to Fig.\ \ref{fig:triangle34} (right), it is consistent with a vanishing cosmological constant, $\Lambda=0$,
which could be the consequence of a dynamical mechanism, such as might arise in quantum gravity
\cite{Coleman:1988tj}.  (The Scalar DM-DE is also 
consistent with $\Lambda=0$, but is ruled out by the 5th force constraint.)

{\bf Noted added.} After submitting this work, we became aware of Ref.\ \cite{Bedroya:2025fwh}, which also studied an exponentially coupled DM-DE model in light of phantom divide crossing and dark fifth force constraints.   Our models are not identical, but they have similar behavior.

\bigskip
{\bf Acknowledgements.}  We thank K.\ Lodha for very helpful information concerning the analysis of Ref.\ \cite{DESI:2025fii}, E.\ Colgain, and D.\ Huterer, M.-X.\ Lin and Tianyi Xie
for enlightening comments; also E.\ Salvioni for pointing out an important correction to the first version, and helpful information on fifth force constraints.  We thank G.\ Levin for his participation during the early stages of this work.  The research of JC and VM was supported by the Natural Sciences and Engineering Research Council (NSERC) of Canada. We also thank Juan Gallego for assistance with the computing cluster.

\bibliographystyle{utphys}
\bibliography{sample}

\vfill\eject

\end{document}